# Genetic studies through the lens of gene networks

This manuscript (permalink) was automatically generated from pivlab/annual_review_of_biomedical_data_science@c71300a on October 11, 2024.

## Authors

- **Marc Subirana-Granés**
  0000-0003-3934-839X · msubirana · msubirana20
  Department of Biomedical Informatics, University of Colorado Anschutz Medical Campus, Aurora, CO, USA

- **Jill Hoffman**
  0000-0002-2690-2593 · jillhoffman
  Department of Biomedical Informatics, University of Colorado Anschutz Medical Campus, Aurora, CO, USA

- **Haoyu Zhang**
  0009-0005-6025-0217 · haoyu-zc
  Department of Biomedical Informatics, University of Colorado Anschutz Medical Campus, Aurora, CO, USA

- **Christina Akirtava**
  0000-0001-7587-2303 · bio2data · bio2data
  Department of Biochemistry and Molecular Genetics, RNA Bioscience Initiative, University of Colorado Anschutz Medical Campus, Aurora, CO, USA

- **Sutanu Nandi**
  0009-0001-5887-999X · snandi-DS
  Department of Pharmacology, University of Colorado Anschutz Medical Campus, Aurora, CO, USA

- **Kevin Fotso**
  0009-0002-2196-6157 · kf-cuanschutz
  Office of Information Technology, University of Colorado Anschutz Medical Campus, Aurora, CO, USA

- **Milton Pividori**
  0000-0002-3035-4403 · miltondp · miltondp · @miltondp@genomic.social
  Department of Biomedical Informatics, University of Colorado School of Medicine, Aurora, CO 80045, USA; Colorado Center for Personalized Medicine, University of Colorado Anschutz Medical Campus, Aurora, CO, USA · Funded by The National Human Genome Research Institute (K99/R00 HG011898); The Eunice Kennedy Shriver National Institute of Child Health and Human Development (R01 HD109765)

✉ — Correspondence possible via GitHub Issues or email to Milton Pividori <milton.pividori@cuanschutz.edu>.

# Abstract

Understanding the genetic basis of complex traits is a longstanding challenge in the field of genomics. Genome-wide association studies (GWAS) have identified thousands of variant-trait associations, but most of these variants are located in non-coding regions, making the link to biological function elusive. While traditional approaches, such as transcriptome-wide association studies (TWAS), have advanced our understanding by linking genetic variants to gene expression, they often overlook gene-gene interactions. Here, we review current approaches to integrate different molecular data, leveraging machine learning methods to identify gene modules based on co-expression and functional relationships. These integrative approaches, like PhenoPLIER, combine TWAS and drug-induced transcriptional profiles to effectively capture biologically meaningful gene networks. This integration provides a context-specific understanding of disease processes while highlighting both core and peripheral genes. These insights pave the way for novel therapeutic targets and enhance the interpretability of genetic studies in personalized medicine.

# Introduction

Understanding the genetics of complex traits remains one of the key challenges in modern genomics. GWAS have significantly advanced our knowledge by identifying associations between genetic variants and traits, shedding light on the genetic architecture underlying a variety of diseases and phenotypic conditions. Since the first GWAS on age-related macular degeneration in 2005, the GWAS catalog has grown to include over 5,000 human traits, encompassing risk loci for conditions such as type 2 diabetes, schizophrenia, and rheumatoid arthritis (1–3). However, most GWAS-identified variants reside in non-coding regions, underscoring the importance of gene regulation in generating phenotypic diversity (4, 5). This gap presents a major challenge in translating GWAS findings into biological insights.

To address this limitation, numerous methods have been developed to link GWAS variants to functional outcomes. Proximity-based methods, such as MAGMA, have been widely used to link GWAS variants to nearby genes based on physical distance (6). However, these methods may not consider long-range acting SNPs, such as the variants in the *FTO* locus that are associated with obesity, which have been shown to influence the *IRX3* gene through long-range interactions (7). While useful, proximity-based approaches may miss causal genes if they are not the closest ones. TWAS has proven effective by linking genetic variants to gene expression, thereby enhancing our ability to identify putatively causal genes for various traits (9). TWAS employs expression quantitative trait loci (eQTL) data to prioritize genes that are influenced by GWAS variants, offering a more mechanistic link between genetic variation and phenotypic changes. Nevertheless, TWAS and similar methods typically focus on individual genes or variants, overlooking the complex interactions among genes that are increasingly recognized as vital to understanding complex traits (10).

The omnigenic model, introduced by Boyle et al. (2017) (10), proposes that the genetic architecture of complex traits involves highly interconnected gene networks. According to this model, core genes directly contribute to a trait, while peripheral genes modulate these core genes through regulatory networks. This conceptual shift emphasizes the need for methodologies capable of capturing polygenic and network-based interactions inherent in complex diseases. Machine learning approaches, particularly those that leverage gene co-expression patterns, are well-suited to this task. Methods such as non-negative matrix factorization and variational autoencoders (VAEs) have demonstrated the ability to extract biologically meaningful gene modules, which represent groups of genes with coordinated expression under specific conditions (12).

In this review, we explore current approaches based on gene modules to integrate genetic studies with other data types. We discuss how integrating machine learning-derived gene modules with genetic and multi-omics data enhances our understanding of complex traits and diseases. We highlight the PhenoPLIER framework ([13](#)), which integrates gene modules derived from transcriptome data with TWAS and drug-induced transcriptional profiles to uncover disease-relevant molecular mechanisms. This approach moves beyond single gene analyses by capturing the broader gene networks that contribute to phenotypic outcomes, offering a more nuanced understanding of the molecular basis of human complex traits and paving the way for more effective, personalized therapeutic strategies.

## Single variant and single gene approaches

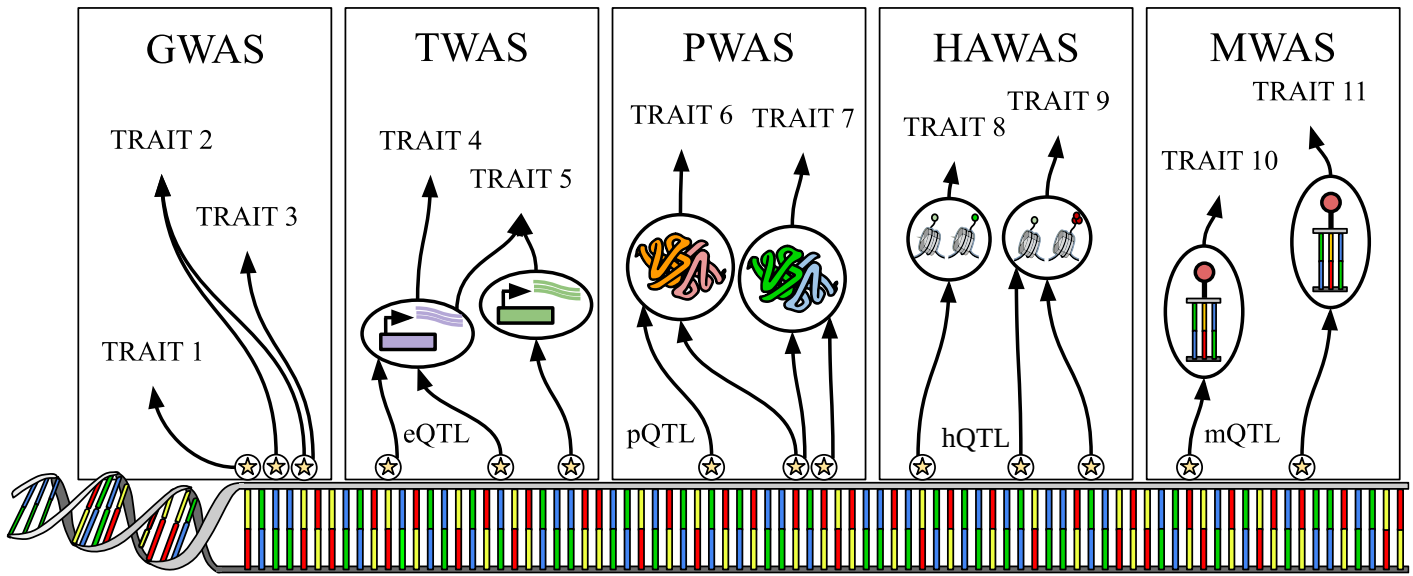

**Figure 1: Major approaches based on single variant and single gene strategies to understand the genetic basis of complex traits**. Genome-wide association studies (GWAS) identify associations between DNA variants and traits. All the other approaches test how different molecular mechanisms might mediate variant-trait associations from GWAS. Transcriptome-wide association studies (TWAS) use variants linked to changes in gene expression (i.e., expression quantitative trait loci or eQTLs). Protein-wide association studies (PWAS) links protein functionality to traits via protein QTLs (pQTLs). Histone acetylome-wide association studies (HAWAS) maps histone acetylation modifications to traits using histone acetylation QTLs (haQTLs), while methylation-wide association studies (MWAS) explores DNA methylation patterns in relation to traits through methylation QTLs (mQTLs).

## From genetic variants to traits: genome-wide association study

In a simple trait, a single gene can be responsible for the disease (monogenic), for example, in the case of Sickle Cell Anemia and Huntington's disease. However, most traits are not so simplistic and are the result of many mutations across the genome (polygenic). Even monogenic traits can sometimes be influenced by the polygenic background ([14](#)). GWAS examine the relationship between specific genetic variants and phenotypes by comparing allele frequencies in individuals of similar ancestry with distinct phenotypic traits (**Figure [1](#)**). Both causal and non-causal variants are found to be significant, and due to the phenomenon of linkage disequilibrium, the results of GWAS are often grouped into risk loci. These loci represent clusters of variants that demonstrate significant associations, although not every variant may be causal ([15](#)).

Since the first GWAS on age-related macular degeneration (AMD) in 2005, the GWAS catalog has rapidly expanded, containing SNP-trait associations across >5,000 human traits ([1](#)). These studies have successfully identified risk loci for a variety of traits such as type 2 diabetes, autoimmune disease, schizophrenia, major depressive disorder, and many more ([2](#), [16](#)). As GWAS participants increase, more loci are able to be identified. For example, a study investigating insomnia with a sample size of >1M identified 202 risk loci (17), compared to earlier studies with a sample size of ~110,000, which were only able to identify 3 risk loci (18). Beyond the identification of risk loci, GWAS have also led to the discovery of novel biological mechanisms in Crohn's disease (19, 20), rheumatoid arthritis (3), and more.

Although GWAS have identified numerous genetic variants associated with complex traits, translating these findings into biological insights remains challenging (2). For example, genes can have epistatic interactions where a secondary locus affects a primary locus. The peripheral effects of the secondary loci are often too small to be picked up through GWAS alone (21). Furthermore, many significant GWAS variants are located in non-coding regions, making it difficult to identify the specific genes that drive trait associations (4). A simple strategy to link GWAS-associated variants with a gene is by physical proximity, which typically selects the closest gene to the most significant SNP at each locus. Approaches like MAGMA (6) are based on proximity to compute a gene-trait association. However, SNPs can have distal effects on the expression of the target gene (5, 7, 22, 23).

Recent approaches combine multi-omics data from various cell types and tissues with GWAS to identify potential mechanisms of SNPs and the associated genes through molecular quantitative trait loci (molQTLs) (24) (Figure 1). For example, an expression quantitative trait locus (eQTL) is a genetic region associated with the expression levels of a nearby or distant gene (25). eQTL studies estimate that the majority of heritability is explained by the combination of multiple weak *trans*-eQTLs (26).

In the next section, we briefly review computational approaches that test different molecular phenotypes via QTLs to link variants with genes. Both GWAS and QTL studies have been penalized by their large multiple testing burden, causing the need to adopt a high level of significance. This results in GWAS being underpowered to detect all heritability explained by SNPs (2).

## From GWAS to gene: transcriptome-wide association studies

TWAS integrates GWAS with gene expression data (**Figure 1**) from eQTL analysis to prioritize genes whose expression across different tissues is influenced by GWAS variants (8, 9). By leveraging predicted gene expression levels, TWAS provides a mechanistic link between genetic variants and traits, allowing researchers to move beyond associations with individual SNPs to identify putatively causal genes (9). Since TWAS models the genetic regulation of gene expression, this approach enables researchers to impute expression levels in GWAS cohorts where expression data may not be available. A key advantage of TWAS over GWAS lies in its ability to increase interpretability by providing a gene-trait association: TWAS connects trait-associated SNPs (which are mostly non-coding) to genes, which are biologically functional units.

Several TWAS approaches have been introduced (27), including PrediXcan (28), FUSION (29), and TIGAR (30). However, all of them implement a similar framework that consists of three steps: 1) model training, 2) gene expression imputation, and 3) gene-trait association. For example, during step 1, PrediXcan builds one expression prediction model per gene and tissue using penalized linear regression with ElasticNet to model sparse genetic architectures. These models contain weights for each SNP used as a predictor for gene expression in a tissue. Given genotype data in a cohort without measured gene expression, during step 2, the SNP weights from the models can be used to impute tissue-specific gene expression for individuals. During step 3, a gene-tissue-trait association is computed by correlating the tissue-specific imputed gene expression with the trait of interest. Most methods (such as Summary-PrediXcan (31) or S-PrediXcan), however, offer a shortcut by computing a gene-tissue-trait association directly from GWAS summary statistics without the need for individual-level data. This process, however, requires the user to select a tissue of interest, which might not be straightforward (9). To address this limitation, approaches such as MultiXcan (32) or UTMOST (33) combine information across tissues to generate a gene-trait association. These multi-tissue approaches are generally more powerful than single-tissue ones, although they do not provide a direction of effect (i.e., whether a higher or lower predicted expression is associated with a higher or lower disease risk).

## Going beyond TWAS

The flexibility of this 3-step framework can also be used to test whether other molecular phenotypes might mediate the association between GWAS variants and a trait of interest. In addition to integrating eQTLs, the framework has been implemented also with protein QTLs (pQTLs), histone acetylation QTLs (haQTLs), methylation QTLs (mQTLs), and splicing QTLs (sQTLs).

PWAS complements GWAS and TWAS by aggregating genetic variations in protein-coding regions to assess their combined impact on protein function and phenotype (**Figure 1**) ([34](#)). Unlike GWAS, which focuses on individual variants, PWAS evaluates the cumulative effects of multiple coding variants, reducing the scale of multiple testing, detecting complex inheritance patterns, and offering deeper insights into potential disease links. PWAS employs the Functional Impact Rating at the Molecular-level (FIRM) machine learning model to evaluate how missense variants affect protein function by assigning impairment scores, while other types of variants are assigned scores using rule-based methods ([34](#)). The aggregate effects of these variants on each gene are then assessed, followed by statistical tests to identify significant associations between gene scores and phenotypes. For example, PWAS confirmed the association of *MUTYH* with colorectal cancer using UK Biobank data, even when no individual variant reached genome-wide significance in GWAS, highlighting its ability to uncover functional associations missed by other methods. However, PWAS relies on high-quality proteomic data, may miss non-coding variant associations, and is computationally intensive.

Epigenome-wide association studies (EWAS) encompass methodologies such as methylome-wide association studies (MWAS) (**Figure 1**) and histone acetylome-wide association studies (HAWAS) (**Figure 1**), based on haQTLs and mQTLs, respectively, which focus on specific types of epigenetic modifications to reveal their roles in gene regulation and disease etiology ([35](#)). MWAS targets DNA methylation patterns, identifying loci where methylation changes are linked to particular traits or diseases. For instance, a study by Shen et al. (2022) demonstrated a causal relationship between DNA methylation and depression, indicating that epigenetic modifications may mediate genetic risk for psychiatric disorders ([36](#)). HAWAS focuses on histone acetylation modifications, which are crucial for regulating chromatin structure and gene expression. Del Rosario et al. (2022) conducted a HAWAS identifying over 2,000 differentially acetylated loci in immune cells from *Mycobacterium tuberculosis*-infected individuals, linking these changes to gene expression and potassium channel genes like *KCNJ15* in modulating apoptosis and Mtb clearance. haQTL analysis further revealed variants associated with immune phenotypes, complementing GWAS findings and enhancing understanding of disease mechanisms ([37](#)). The tissue-specific nature of epigenetic modifications, alongside the capacity to capture cellular plasticity and environmental influences, enhances our insight into the effects of genetic variants across distinct tissues, temporal cellular states, and gene-environment interactions ([38](#), [39](#)).

Despite the advancements facilitated by single variant and single gene methodologies, a common thread persists: the focus remains on one gene at a time. The underlying expectation is that identifying a gene linked to a trait will directly unveil the biological mechanisms driving disease processes. While this has been successful in certain monogenic disorders, complex traits often involve intricate interactions among multiple genes and environmental factors. Consequently, the one-gene-at-a-time paradigm may oversimplify the multifaceted nature of these traits. In light of these considerations, it becomes imperative to reevaluate our approaches to dissecting the genetic architecture of complex traits. This necessitates a shift towards methodologies that can capture the polygenic and network-based interactions inherent in complex diseases.

# From single genes to gene networks: the omnigenic model for complex traits

## Omnigenic model as a framework for deciphering genetic architecture of traits

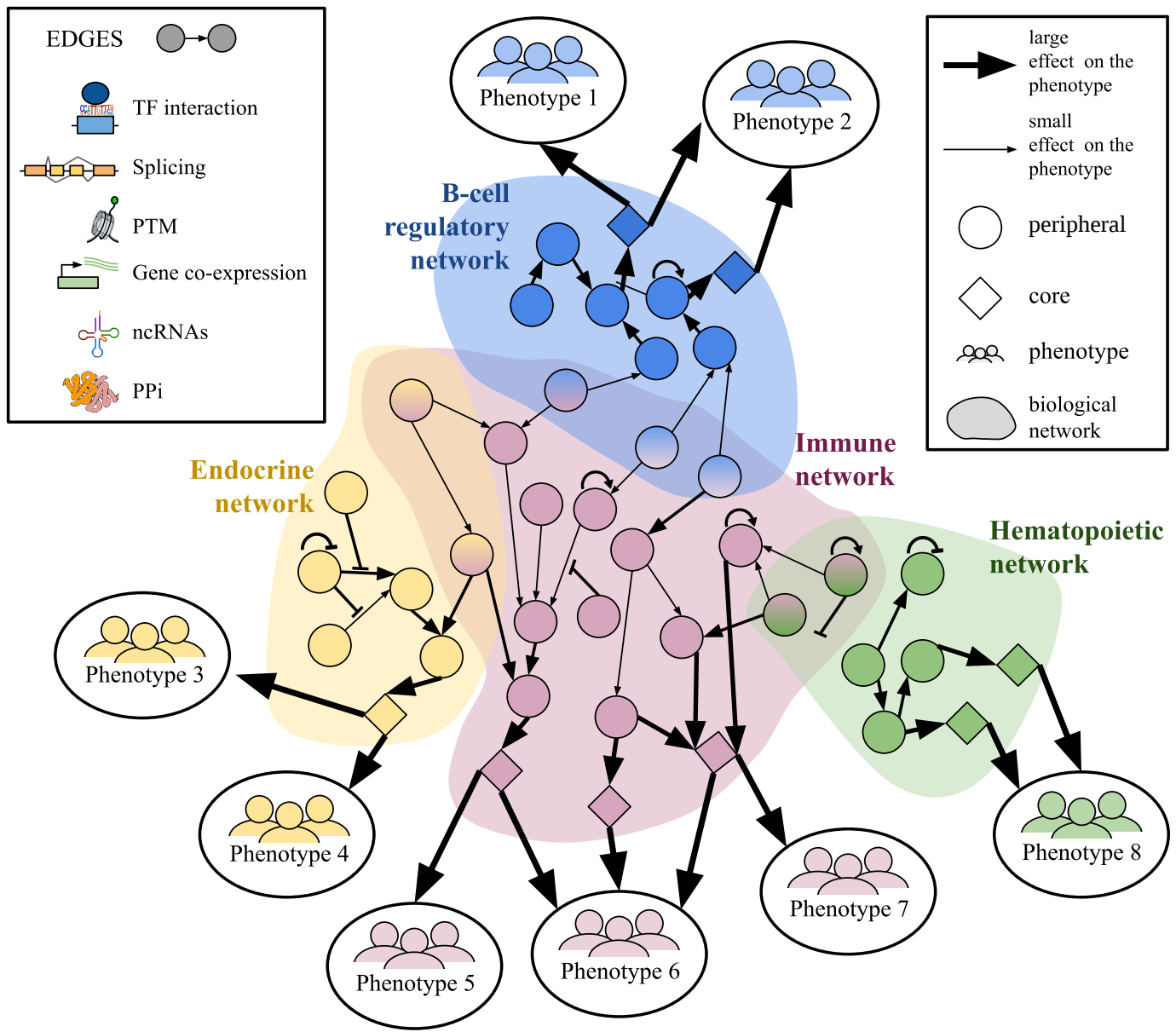

**Figure 2: Schematic of the omnigenic model.** Schematic representation of the biological networks and associated phenotypes in an omnigenic context. Nodes represent individual genes, while edges indicate functional relationships between the nodes, such as transcription factor (TF) binding, splicing events, post-translational modifications (PTMs), gene co-expression, non-coding RNA (ncRNA), and protein-protein interactions (PPIs). The size of the arrows connecting phenotypes indicates the magnitude of their effect, with thicker arrows representing larger phenotypic influences. iological networks are indicated by shaded regions, where genes might play a role in different networks. Genes are categorized as "core" (diamonds) if they have a direct effect on the phenotype, or as "peripheral" (circles) if they have an indirect effect on the phenotype by moduling core genes. Phenotypes are linked to their relevant biological networks, highlighting the interplay between different systems and their contribution to phenotypic outcomes.

In the early 20th century, a significant debate arose between Mendelian geneticists and biometricians concerning the inheritance of complex traits. Mendelians emphasized single gene traits with discrete inheritance patterns, while biometricians argued that such models could not account for the continuous phenotypic variation observed in populations. This conflict highlighted a fundamental question: how can discrete genetic factors produce continuous phenotypic variation? ([10](#), [40](#)). The disagreement was resolved by R.A. Fisher in 1918, who introduced the infinitesimal model, demonstrating that if many genes influence a trait, their combined effects can generate continuous phenotypic distributions ([41](#)). Despite the success of Fisher's infinitesimal model, the actual number of genes involved in complex traits and the magnitude of their effects remained uncertain for much of the 20th century. Translating genotypes into phenotypes proved challenging due to the complexity of the genetic mechanisms driving trait variation. In this context, GWAS brought a new perspective for identifying specific genetic variants with functional effects on complex traits. Over a decade of GWAS has led to several unexpected findings:

1. The strongest GWAS associations exhibit modest effect sizes on disease risk, collectively accounting for only a minor fraction of the predicted heritability (missing heritability) ([42](#)). It is now understood that much of this missing heritability arises from a vast number of common variants with small effects that remain undetected in current sample sizes ([43](#)).

2. A surprising uniform distribution of these small-effect significant variants exists throughout the genome, including non-coding regions, with a notable concentration in regulatory elements such as enhancers and promoters ([44](#), [45](#)). For instance, it is estimated that 71%–100% of 1-MB windows in the genome contribute to heritability for schizophrenia ([46](#)).

3. Trait variation is influenced by the broader genome, not only by genes with direct biological links to specific phenotypes. However, the idea that nearly all genomic variants could exert direct additive effects on a given phenotype is biologically unlikely ([10](#)).

In response to these unexpected findings, Boyle et al. (2017) proposed the omnigenic model ([10](#)), a conceptual framework in which the genetic architecture of complex traits could be explained by highly interconnected gene regulatory networks. The omnigenic model reframes the earlier infinitesimal model by distinguishing between "core" genes, which have a direct biological role in the disease or trait, and "peripheral" genes, which influence core genes and thus indirectly affect the disease or trait through regulatory networks ([10](#)). The omnigenic model, unlike the traditional infinitesimal model, offers a mechanistic rationale for complex trait architecture based on molecular and cellular biology, understanding the molecular pathways that connect genetic variation to phenotypic traits (**Figure [2](#)**).

The key proposals of this model are: 1) nearly all genes expressed in cells relevant to the trait have the potential to impact the regulation of core genes, and 2) for typical traits, nearly all heritability is attributable to variation near peripheral genes. Consequently, while core genes function as the primary drivers of disease, it is the cumulative effects of numerous peripheral gene variants that determine polygenic risk ([47](#)). This framework provides a possible explanation for several previously identified problems, including widespread pleiotropy (the ability of a single gene to affect multiple traits), polygenicity (the involvement of many genes in the manifestation of a single trait), the bias in effect size where small effect variants can collectively account for a large portion of heritability, and the uniform distribution of these effects across the genome.

## Success stories of the omnigenic model

Since the introduction of the omnigenic model, numerous studies have adopted it as a framework to deepen our understanding of the genetic architecture of complex traits and the molecular mechanisms that drive them.

In schizophrenia, the application of the omnigenic model has revealed specific core gene sets associated with the disorder. Notably, the *TCF4* gene set maintained its significant impact even after excluding SNPs within the *TCF4* gene itself. This finding suggests that peripheral genes within the set contribute to the development of schizophrenia. The omnigenic model has also helped propose potential causes for disorders like autism (48). Additionally, the observed systematic floor effect across polygenic scores aligns with the model's prediction that most genes expressed in relevant cells contribute to heritability, thereby highlighting the model's potential in explaining the broad genetic contributions to complex psychiatric conditions (49).

Li et al. (2024) identified a significantly connected subgraph formed by cancer-affected coding genes and ncRNAs by focusing on connectivity as a key topological feature. This approach highlighted the essential role of ncRNAs in linking fragmented cancer-affected genes, consistent with the omnigenic framework's premise that peripheral genes, including non-coding elements, contribute to complex traits such as cancer. Importantly, the inclusion of ncRNAs enhanced the identification of cancer-related pathways, indicating that a comprehensive network model encompassing ncRNAs is more effective in characterizing disease relationships than models concentrating solely on coding genes (50).

Empirical support for the omnigenic theory extends beyond human diseases. Chateigner et al. (2020) demonstrated this concept in European black poplars, showing that both core and peripheral genes play a crucial role in predicting phenotypes. While core genes are indeed important, the information they provide must be complemented by other genes to ensure accurate phenotype predictions. Furthermore, peripheral genes were found to carry significant biological information, contributing to robust predictions (51).

## Limitations of the omnigenic model: from a theoretical to a practical approach

In essence, the omnigenic model proposes a specific mechanism: the genetic architecture of complex traits is intrinsically interconnected, involving a multitude of genes that contribute to phenotypic variation both directly (core genes) and indirectly (peripheral genes). This framework facilitates an understanding of the highly polygenic and often subtle genetic effects on complex traits, shifting the focus from individual genes to gene networks. While this approach has proven successful in some fields, the model exhibits certain limitations and has faced several criticisms.

One major concern is the lack of clear, quantitative methods for identifying peripheral genes or estimating their contribution to heritability. While the model suggests that most genetic effects arise from peripheral genes, it remains unclear how to statistically define and measure these effects (52, 53). Interestingly, research suggests that most heritability in complex traits is controlled by *trans* effects rather than *cis* effects, despite their small effect sizes (47). This paradox can be explained by natural selection acting most strongly against variants with large effects, thereby limiting the contribution of directly biologically important genes to heritability (47). Thus, understanding *trans* effects, where a variant at one locus influences genes at distant loci, is crucial for identifying and quantifying peripheral genes and their contribution to core traits. However, detecting and quantifying *trans* effects remains a significant challenge due to their weak individual effects.

The omnigenic model postulates that nearly all genes expressed in cells relevant to a given trait have the potential to influence the regulation of core genes. However, Yengo et al. (2022), in their recent GWAS on height, challenge this view. They introduced the first "saturated" GWAS, where further increases in sample size are unlikely to yield additional genetic insights unless participant diversity or variant inclusion is expanded. Their findings revealed that only a subset of genomic regions (~20%) seems to contribute to height determination in individuals of European ancestry. This suggests that, at least for height, not all genes play a role in the trait as predicted by the omnigenic model (54).

Furthermore, the omnigenic model has been criticized for its binary classification of genes into core and peripheral categories, which might oversimplify biological systems and potentially underestimate their true complexity (52). It also fails to account for gene-environment interactions, which play a crucial role in shaping complex traits (55). Additionally, while the model provides a conceptual framework, it remains unclear how to translate it into a practical statistical model (56).

As George E. P. Box remarked, "all models are wrong, but some are useful." While we acknowledge that the omnigenic model simplifies the inherent complexity of biological systems and may not universally apply to all traits, we maintain that it remains a valuable framework for elucidating genetic architectures. The model effectively bridges quantitative and molecular genetics, offering comprehensive mechanistic insights with predictions on quantitative variation. Moreover, the omnigenic model has been instrumental in shifting the focus from single gene analyses to network-based models. Recognizing the necessity of translating this model into practical applications, machine learning-derived gene modules, which use this concept to infer gene-gene networks, offer a promising way to ground these theoretical ideas into practical approaches.

# From gene networks to machine learning derived gene modules: hands-on strategies for inferring gene-gene interactions

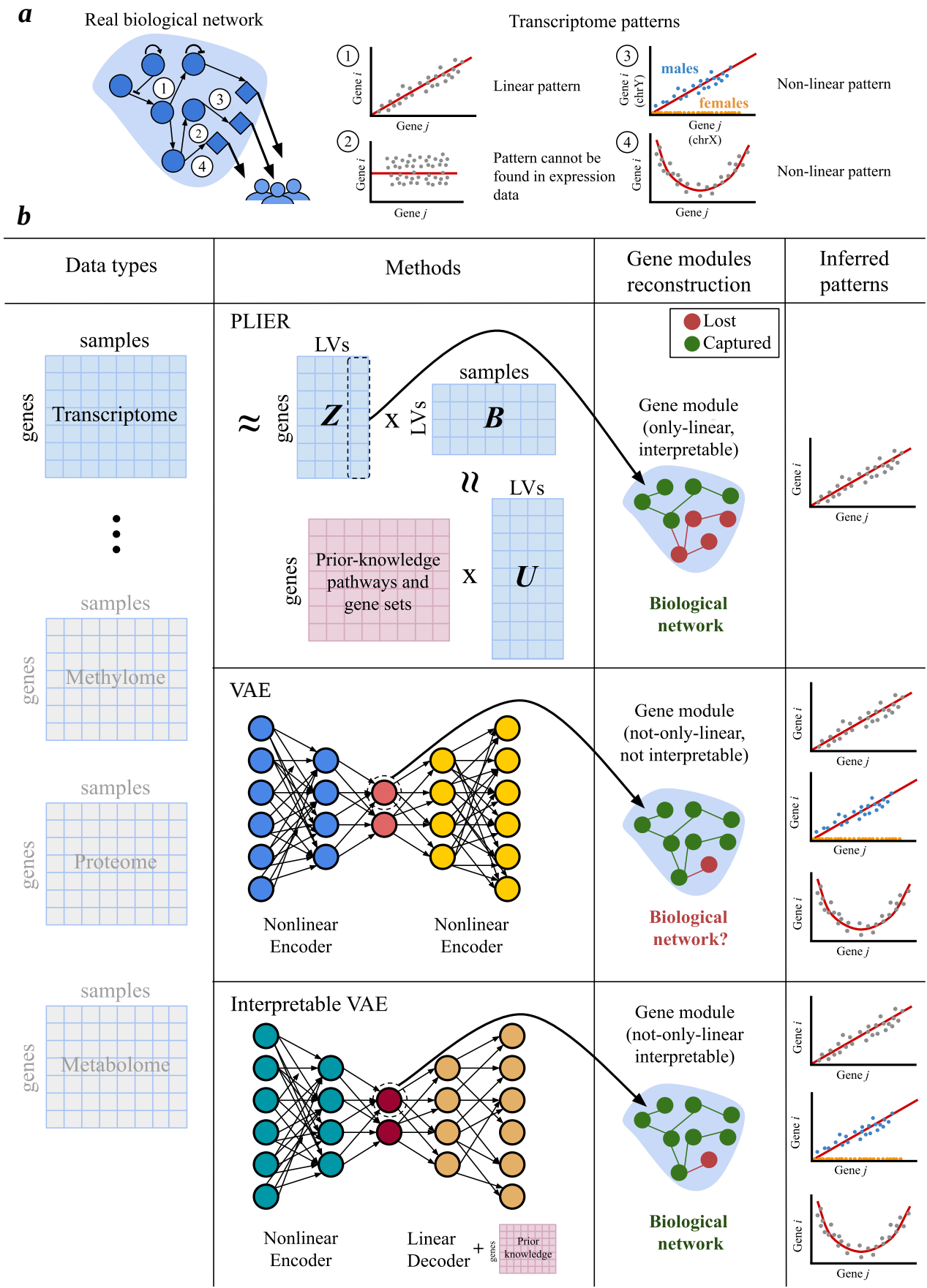

**Figure 3: Computational methods for machine learning derived gene module detection.** *(a)* Real biological network, where nodes represent genes and edges denote the relationships between them. It captures all possible patterns that can link genes. Some of these relationships are gene co-expression patterns, which transcriptomics can capture, such as (1), (3) and (4). Other patterns, such as protein-protein interactions (2), are not captured by transcriptomics and are thus not reflected in the co-expression analysis. *(b)* Different approaches for capturing machine learning-derived gene modules. In this example, transcriptome data is used as input, but other data types such as methylomics, proteomics, and metabolomics can also be applied. The first row illustrates the Pathway-Level Information Extractor (PLIER) method, which uses non-negative matrix factorization and prior knowledge to align gene modules with known biological pathways. This approach generates interpretable gene modules but only captures linear relationships. The second row shows a variational autoencoders (VAEs), which can capture both linear and nonlinear patterns but does not integrate prior knowledge, resulting in less interpretable gene modules. The third row highlights interpretable VAEs, which incorporate prior knowledge into its decoder, achieving a balance between capturing complex relationships and maintaining biological interpretability.

# From single gene analysis to gene module discovery with machine learning approaches

High-throughput technologies, particularly genome-wide gene expression profiling tools such as RNA sequencing (RNA-seq), have fundamentally transformed the landscape of molecular biology. Unlike traditional gene-by-gene approaches, RNA-seq offers a comprehensive view of the transcriptomic landscape ([57](#)). These advancements, alongside conceptual frameworks like the omnigenic model, have facilitated a paradigm shift from a single gene perspective to a module-based approach, wherein groups of genes, rather than individual genes, are essential for elucidating the complexities of biological networks ([58](#)).

Gene modules, a key type of biological network, consist of nodes (genes) and edges that reflect coexpression relationships between them (**Figure [3](#)*a***). These modules, formed by genes with coordinated expression patterns under specific biological conditions, not only reveal coexpression but could also provide valuable insights into regulatory mechanisms, gene functions, and the pathways involved in traits and diseases ([59](#)).

Numerous approaches have been proposed for module detection in gene expression data. Machine learning techniques are particularly proficient at addressing common challenges in omics data, such as data sparsity and high dimensionality ([60](#)). Machine learning methods can compress correlations into lower-dimensional representations, facilitating the processing of large datasets to identify intricate coexpression patterns. Additionally, machine learning approaches are capable of integrating various types of omics data, enhancing the detection of gene modules and providing a more comprehensive understanding of biological networks ([60](#)).

In this context, unsupervised machine learning approaches have gained significant popularity ([61](#)). These methods facilitate the discovery of patterns and structures within the data without requiring prior knowledge of the system, allowing for the inference of novel findings and hypotheses. This capability is particularly advantageous for gene module inference, which aims to identify groups of co-expressed and potentially co-regulated genes that contribute to similar biological functions ([62](#)).

The most common approach in unsupervised machine learning is clustering, with methods such as WGCNA, which has been widely used since the first gene expression datasets became available and remains one of the most popular tools today ([62](#)). However, these methodologies typically rely on traditional correlation measures (e.g., the Pearson correlation coefficient), which are limited to capturing linear associations between continuous data. Newer approaches, such as the Clustermatch Correlation Coefficient (CCC), have been developed to quantify both linear and nonlinear correlations in complex gene expression datasets, providing a more flexible means of detecting intricate relationships that are not apparent through linear models alone ([63](#)).

In this section, given their successful applications, we will focus on decomposition approaches like Principal Component Analysis (PCA) and deep learning (DL) methods for inferring gene modules, highlighting their effectiveness in uncovering complex gene interactions and enhancing our understanding of biological networks (**Figure [3](#)*b***). However, we acknowledge that utilizing transcriptional data alone is insufficient for fully comprehending the intricate system of biological networks; it serves as an initial step toward a more comprehensive understanding.

## Decomposition methods

Decomposition methods reduce complex, high-dimensional gene expression data into simpler components, revealing underlying patterns and structures that aid in interpreting large-scale datasets. These methods outperform other unsupervised machine learning approaches in recovering gene modules by effectively capturing local co-expression effects present in only a subset of biological samples, while also allowing genes to be assigned to multiple modules ([62](#)).

Simple models like PCA have traditionally been used to reduce dimensionality in gene expression data, offering a basic understanding of co-expressed genes ([64](#)). However, these models often lack the ability to incorporate prior biological knowledge, which limits their interpretability and biological relevance. To overcome these limitations, more sophisticated semi-supervised approaches such as the Pathway-Level Information Extractor (PLIER) ([11](#)) or GenomicSuperSignature ([65](#)) have been developed. These methodologies combine unsupervised approaches (decomposition methodologies) with supervised approaches to infer gene modules and annotate them with prior biological knowledge.

GenomicSuperSignature employs PCA to identify gene modules applying prior knowledge for biological annotation after module discovery. This flexibility allows it to adapt to novel datasets and provides computational efficiency even with large datasets. However, the lack of prior knowledge integration during training can result in modules that are less biologically interpretable compared to those from PLIER-based methods ([65](#)).

PLIER (**Figure [3](#)*b***) is a semi-supervised framework based on non-negative matrix factorization. During training, it reduces high-dimensional gene expression data into a smaller set of latent variables (LVs) while aligning some of them to known pathway or gene set annotations. These LVs are designed to capture the greatest variance in the data, with a fraction of the LVs representing known mechanisms (i.e., aligned with prior knowledge/pathways), while others capture potential technical artifacts or novel patterns. By incorporating prior pathway/gene set information, PLIER generates interpretable latent representations. Importantly, PLIER retains some gene modules that are not aligned with pathways, allowing researchers to distinguish LVs that represent either known patterns, technical artifacts, or potential novel biological insights ([55](#)).

PLIER has been used to integrate large-scale gene expression data using transfer learning across multiple cell types and conditions ([66](#)). By leveraging large datasets like recount2 ([67](#)), it enhances the accuracy of inferred gene modules by identifying patterns consistent across diverse data. This application addresses the limitation of small datasets, such as those for rare diseases, by training a PLIER model on a large public data compendium and applying it to smaller datasets, resulting in models that align well with biological factors. This approach improves the interpretability and reliability of gene expression patterns, particularly in complex experimental contexts.

PLIER can be used in specific cohort contexts, focusing on dissecting traits within smaller, more homogeneous datasets. Nandi et al. (2024) demonstrated its utility in the Human Trisome Project (HTP) ([68](#)), identifying gene modules that mediate the effect of karyotype on body mass index in Down syndrome, highlighting key regulators such as *GPX1* and *MCL1*. This allows it to capture condition-specific regulatory mechanisms, but at the expense of broader generalizability, making it a powerful tool for understanding unique biological processes within specialized groups.

PLIER, initially designed for human gene expression data, has been successfully adapted for use in other species, demonstrating its versatility across different biological contexts. By applying the framework to model organisms, researchers can infer gene modules in species beyond humans, expanding the utility of PLIER for comparative studies and translational research. One such adaptation is MousiPLIER, which tailors the PLIER framework for use in mouse models, a widely used system in biomedical research ([69](#)).

## Deep learning methods: autoencoders

In contrast to linear methods, non-linear models such as DL approaches excel at capturing complex patterns in data. This strength has been demonstrated in various fields, including predicting gene expression directly from sequence data ([70](#)–[72](#)).

DL models are also highly effective in unsupervised settings, such as inferring gene modules. Specifically, DL methods like autoencoders ([73](#)), which utilize neural networks to learn compressed representations, can capture both linear and nonlinear patterns between genes. This makes autoencoders particularly suited for uncovering gene modules and efficiently handling high-dimensional datasets. For instance, VAEs (**Figure [3](#)*b***) enhance this process by encoding data into a continuous latent space, enabling dimensionality reduction and the generation of new data. However, this increased power to capture more general patterns has a major trade-off: interpretability. To overcome this, interpretable VAE models like the Pathway Module VAE (pmVAE) ([74](#)), expiMap ([75](#)), and VEGA ([12](#)) integrate biological pathway information into the decoder, which improves interpretability, thus making these approaches function as linear dimensionality reduction methods (**Figure [3](#)*b***).

From now on, we refer to a gene module as a pattern extracted by any unsupervised approach (such as PLIER, GenomicSuperSignature, VAE, clustering, etc.) applied to gene expression data, which could be a latent variable or other types of gene cluster.

## A gene module perspective for genetic studies

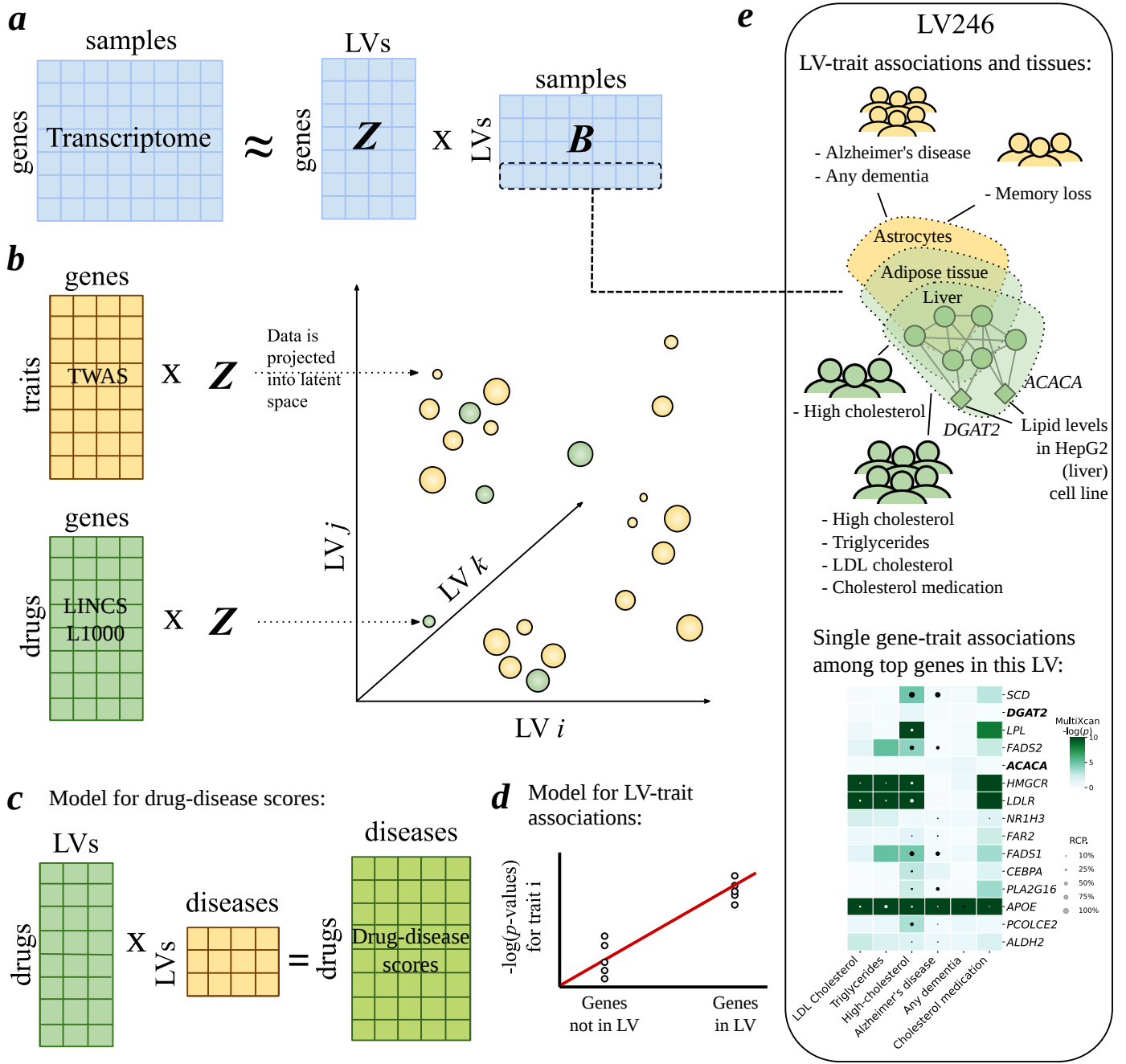

**Figure 4: An integrative, gene module-based approach for genetic studies.** Each panel shows a component of the PhenoPLIER framework (13). *(a)* First, latent variables (LVs) or gene modules are learned from transcriptome data using the Pathway-Level Information Extractor (PLIER) matrix factorization method. PLIER generates matrix $\mathbf{Z}$, which has gene weights for each module, and matrix $\mathbf{B}$, which has the samples in the latent space. *(b)* Schematic of gene-trait associations from Transcriptome-wide association studies (TWAS) and gene-drug scores from LINCS L1000 being projected into the latent space for a joint analysis. *(c)* Schematic of a gene module-based drug reporposing framework, where the projection of TWAS and LINCS L1000 data is used to compute a drug-disease score. *(d)* Schematic of a regression model that tests whether genes that belong to a module (using a column of $\mathbf{Z}$) tend to be more strongly associated with a trait (using $p$-values from TWAS). *(e)* (top) Example of a gene module identified as LV246 analyzed in (13). *DGAT2* and *ACACA*, found in a CRISPR screen to be linked to lipid metabolism and potential core genes, were among the top 15 genes in this module. Using metadata from the top samples for LV246 from $\mathbf{B}$, the module was found to be expressed mainly in adipose tissue, and liver and astrocyes were part of the top 10 cell types and tissues. Using the regression model depicted in d), this module was significantly associated with high cholesterol, tryglycerides, LDL cholesterol, cholesterol lowering medication, Alzheimer's disease and dementia in the discovery cohort (UK Biobank), and high cholesterol and memory loss in the replication cohort (eMERGE). (bottom) The $p$-values and colocalization probability from gene-trait associations from TWAS for a subset of the top 1% genes in LV246 ($y$-axis) and traits significantly associated with LV246 ($x$-axis). It can be seen that *DGAT2* and *ACACA* (in boldface), although strongly linked to lipid metabolism, are weakly associated with these lipid-relevant traits.

All the previously described approaches provide valuable and orthogonal information to better understand the molecular basis of human complex traits, including different molecular mechanisms through variant-trait, variant-gene, gene-trait, and gene-gene information. All of them have limitations and offer a partial view of the complex molecular interplay underlying pathophysiological processes. In this section, we describe a gene module-based approach aimed at identifying patterns shared across these data sources. The underlying hypothesis of this approach is that biologically meaningful, disease-relevant gene-gene links will be revealed when different yet incomplete data modalities are integrated. In particular, we show examples of how the integration of gene modules with gene-trait associations and gene-drug data highlights core genes, disease-relevant molecular processes, and drugs' mechanisms of action.

The PhenoPLIER framework ([13](#)) is a recent computational framework that implements this gene module-based approach. PhenoPLIER can integrate gene modules extracted from any transcriptome data with gene-trait associations from TWAS and drug-induced transcriptional profiles (**Figure [4](#)**). The approach integrates different molecular data sources by using a latent space derived from transcriptome data where each LV is a gene module (**Figure [4](#)a**). Then, different data modalities are integrated into this common latent space for a joint analysis, which has been shown to detect important genes missed by standard methods, capture relevant trait clusters, and predict drug-disease links more accurately than state-of-the-art drug repurposing approaches. An advantage of PhenoPLIER is interpretability: it uses gene module models from PLIER (**Figure [4](#)a**), which provide not only information about which genes belong to a module but also what the top samples are where those genes are expressed. If present, by looking at the metadata of the RNA-seq datasets, it is possible to infer if those samples represent consistent conditions such as tissues, cell types, or other more complex contexts.

In PhenoPLIER, this latent transcriptional space is described by 987 gene modules derived from recount2 ([66](#), [67](#)) (**Figure [4](#)a**), a large and highly heterogeneous expression compendium. Given the heterogeneity of the dataset, these gene modules can capture genes expressed across different contexts such as tissues, cell types (across differentiation stages), and different disease states or stimuli. These contexts can be extracted from sample metadata. PhenoPLIER incorporates gene-trait associations computed with the PrediXcan family of TWAS methods ([28](#), [31](#), [32](#)) on two different cohorts: the UK Biobank ([76](#), [77](#)) as discovery, and the Electronic Medical Records and Genomics (eMERGE) network phase III ([78](#), [79](#)) as replication. Gene-trait associations are then projected into the latent space (**Figure [4](#)b**), where cluster analysis on traits was performed, and expected and stable groupings of traits were detected, such as asthma and allergies, heel bone-densitometry measurements, hematological assays on red blood cells, physical measures, keratometry measurements, assays on white blood cells and platelets, skin and hair color traits, autoimmune disorders, and cardiovascular diseases. The cardiovascular grouping also included other cardiovascular-related traits such as hand-grip strength ([80](#)), and environmental/behavioral factors such as physical activity and diet. The PhenoPLIER approach also proposed a methodology based on interpretable classifiers to detect which gene modules are driving the different clusters. For example, gene modules associated with the red blood cells cluster were 1) well-aligned to pathways related to early progenitors of the erythrocytes lineage, 2) predominantly expressed in early differentiation stages of erythropoiesis, and 3) strongly associated with different assays on red blood cells from TWAS. Other gene modules were associated with the keratometry measures grouping and expressed in corneal endothelial cells, and the grouping with autoimmune disorders was driven by gene modules expressed in T cells. These results show that shared patterns exist between prior knowledge (pathways), gene expression, and gene-trait associations from TWAS.

In PhenoPLIER, information about gene-drug links is incorporated from transcriptional responses to small molecule perturbations profiled in LINCS L1000 ([81](#)). This information was integrated with gene-trait associations to build a gene module-based drug repurposing framework with the purpose of assessing how substituting LVs for single genes predicted known treatment-disease relationships better (**Figure 4***c*). Based on an established drug repurposing strategy that matches reversed transcriptome patterns between genes and drug-induced perturbations (82, 83), the authors adapted an existing drug repurposing framework (84) that uses single gene information from TWAS to gene modules in PhenoPLIER. Then, these two approaches, the single gene-based and gene module-based, were compared using a manually curated gold standard set of drug-disease medical indications (85, 86) for 322 drugs across 53 diseases to evaluate the prediction performance. The gene module-based approach outperformed the single gene-based one with an area under the curve of 0.63 vs. 0.57, and average precision of 0.86 vs. 0.64. Although the performance difference in this task was not large, the authors noted that the gene module-based approach represents a compressed version of the entire set of single gene-based results, and the higher performance implied that the low-dimensional latent space used (which necessarily misses some information) captured biologically meaningful gene-gene patterns. Additionally, since gene modules represent interpretable features, the authors found that lipid-related gene modules expressed in adipose tissue and liver were among the top modules contributing to the prediction of high cholesterol and Nicotinic acid (Niacin, which can treat lipid disorders), potentially resembling known mechanisms of action of Niacin, such as decreasing the production of low-density lipoproteins (LDL) either by modulating triglyceride synthesis in hepatocytes or by inhibiting adipocyte triglyceride lipolysis (87).

The PhenoPLIER framework is also able to compute an association between a gene module and a trait by integrating single gene TWAS results (**Figure 4***d*). The association is computed using a regression model that tests whether genes that strongly belong to a module (using a column in matrix Z in **Figure 4***a*) are also strongly associated with the trait (using a row in the yellow TWAS matrix in **Figure 4***b*). For validation, the study conducted a CRISPR-Cas9 screen in the HepG2 (liver) cell line to identify genes associated with lipid regulation and found a high-confidence lipid-increasing gene set that included genes *DGAT2* and *ACACA*, which fit the definition of core genes. The gene module identified as LV246 (**Figure 4***e*) was the module most strongly enriched with this lipid-increasing gene set, containing *DGAT2* and *ACACA* among the top 15 genes most strongly connected with the module. LV246 was found to be 1) aligned with lipid metabolism pathways, 2) expressed in adipose tissue, liver, and astrocytes, among others, and 3) significantly associated with lipid-related traits in the discovery cohort (UK Biobank; **Figure 4***e*) and replication cohort (eMERGE), such as high cholesterol, triglycerides, LDL cholesterol, and cholesterol-lowering medications in the UK Biobank, and high cholesterol (hyperlipidemia, phecode 272.11) in eMERGE. Although a direct link between lipids and *DGAT2* and *ACACA* was found by the CRISPR screening, the authors noted that these genes lack strong TWAS associations with any of these lipid-related traits (**Figure 4***e*, bottom). The GWAS Catalog shows that genetic variants in *DGAT2* were found to be associated with cholesterol levels (88), but *ACACA* lacks any association with these phenotypes (89). A potential explanation provided by the study suggests that, based on the omnigenic model, *DGAT2* and *ACACA* could represent core genes, and most of the other genes in LV246 might be peripheral genes that potentially *trans* regulate them. This example suggests that strong GWAS hits could represent either peripheral or core genes, and that some core genes might be missed by GWAS alone.

Additionally, another interesting trait association in module LV246 was with Alzheimer's disease (AD) and dementia in the UK Biobank, as well as memory loss (phecode 292.3) in eMERGE, which tends to worsen as AD progresses (90). The association with AD is relevant since this module was also found to be expressed to some degree in astrocytes and has *APOE* as one of the top genes (**Figure 4***e*, bottom). *APOE*, the gene most strongly associated with late-onset AD (91), encodes apolipoprotein E (ApoE), which is involved in lipid transport in the brain (92), a mechanism that, when dysregulated, may play an important role in AD pathogenesis. From TWAS alone, it can be seen that *APOE* is strongly associated and colocalized with the same traits as LV246 (**Figure 4***e*, bottom). However, a recent systematic survey across clinical trials that evaluate ApoE-targeted drugs for AD found modest to no efficacy in the treatment for this disease (93). Although there is strong evidence that *APOE* is a causal gene for AD, it still remains to be determined whether it represents a core gene or plays a more peripheral role. A gene module-based approach might help prioritize core genes that remain elusive when using standard single gene strategies.

Another approach to compute an association between gene modules and traits is MAGMA gene-set analysis (6). In fact, the regression model employed by PhenoPLIER is based on MAGMA. The difference is that PhenoPLIER uses gene-trait association from the PrediXcan family of TWAS methods, while MAGMA does not incorporate eQTL data and uses a proximity-based approach in linking variants with genes. Another difference is that MAGMA allows for conditional and interaction gene-set analysis to account for correlated gene modules, and this approach was found to be useful in detecting novel pathways in the context of blood pressure (94). The use of eQTL data by TWAS, however, might better fit the gene regulatory network assumptions of the omnigenic model than proximity-based approaches like MAGMA.

Approaches based on gene modules have the potential to go beyond prior knowledge and capture patterns that might be unique to different disease datasets. First, although a gene module can be aligned with a pathway, this does not mean that it is restricted to prior knowledge as in standard pathway analyses. Being aligned with pathways means that a module resembles a known mechanism, which helps in interpretability and in distinguishing it from other patterns that might be related to technical noise. However, the module could also capture other genes that are not part of the pathway but potentially involved in the function. Second, gene modules can be extracted from very large, heterogeneous datasets such as recount2, or more specific ones such as the Human Trisome Project, which includes gene expression data from people with Down Syndrome (DS). Approaches such as PLIER and PhenoPLIER have been fundamental in extracting DS-specific gene modules related to obesity, a common co-occurring condition (95).

A disadvantage of the PhenoPLIER approach is that gene modules are generated by an algorithm, and as such, they could represent artifacts or be aligned with technical noise. Interpretable methods such as PLIER and some VAE models that help segregate technical noise from relevant biology are key to solving this problem. Another limitation is the potential difficulty in determining the contexts in which a gene module is expressed. In the example of LV256 and lipid-related traits, the module was also found to be expressed, to varying degrees, in tumor samples, skin, skeletal muscle, and other contexts that might be hard to interpret. The interpretability advantage is also limited by the quality of the RNA-seq metadata. Additionally, compared with GWAS and TWAS, where the unit is an objective molecular entity (a SNP or gene in a particular chromosome and position), a gene module not only might or might not exist, but the algorithm could fail to capture all gene modules relevant for a particular study, which could bias any downstream analysis. Finally, a gene module, equivalent to a gene co-expression network, represents a set of correlated genes' expression in some conditions, but it does not provide information about gene-gene links (it's not directed) and could include many false positives (96). Future approaches should also incorporate other data modalities to refine a gene module, such as transcription factor binding data or chromatin accessibility.

Although there are several limitations when we only correlate gene pairs across RNA-seq samples (96), integrating gene networks with other data modalities has the potential to identify disease-relevant molecular mechanisms (55). Here, we describe an approach that integrates gene modules with other data sources, such as gene-trait and gene-drug information. The underlying hypothesis of this approach is that biologically meaningful gene-gene links will be captured when different data modalities that at least partially capture these links are integrated. In our examples, these data modalities are gene modules from transcriptome data, single gene-trait associations, and single gene-drug links via drug-induced transcriptional profiles. We also show that if the unsupervised approach used to extract a gene module is also interpretable (i.e., provides information about the specific contexts, such as cell types or tissues, that explain why those genes were grouped together), the data integration approach also sheds light on potential context-specific transcriptional mechanisms.

## Conclusions and future perspectives

Genomics plays an important role in realizing the promise of personalized medicine, although not without significant challenges. In this review, we discuss different molecular phenotypes, theoretical models, and computational approaches that offer various mechanistic views on how genetic variation leads to phenotypic changes. To fully understand the molecular basis of human complex traits, it is essential to integrate diverse datasets and unravel the complex structures of biological networks. Gene co-expression networks, represented as gene modules, have been successfully used across a variety of fields ([97](#)–[99](#)). Here, we review the main components of a gene module-based approach that leverages gene co-expression patterns learned from large transcriptomic datasets to integrate GWAS, TWAS, and drug-induced transcriptional profiles. Drawing on modern theories of the genetic architecture of complex traits, we show that these approaches can prioritize core genes and create interpretable frameworks for drug repurposing.

Gene-gene interactions are a key component in the omnigenic model. While the integration of gene modules with gene-trait associations and gene-drug data has been a valuable first step, it represents only a fraction of the complexity inherent in biological systems. Biological networks extend far beyond these initial layers, involving intricate, multilayered systems such as histone modifications, transcription factor binding sites, *cis*-regulatory networks, and chromatin conformation—all of which contribute to the co-regulation of gene transcription. Moreover, this complexity includes protein-protein interaction networks, which form the backbone of signaling pathways and facilitate the transmission of signals both within and between cells. Additionally, metabolic networks influence enzyme activities and metabolite levels, propagating changes that affect numerous other molecules ([55](#), [100](#)). Incorporating and quantifying these additional biological layers will enhance gene module construction, offering a more comprehensive and accurate representation of the intricate nature of real biological networks.

A gene module approach that integrates genetic studies has several important components: 1) gene expression data relevant to the research problem, such as large heterogeneous datasets that span different conditions, or otherwise focused, smaller datasets that capture more subtle, unique, and disease-specific gene modules; 2) unbiased and relevant prior information that can be used during or after module extraction to distinguish relevant transcriptomic signatures from technical noise; and 3) reliable single variant/gene associations with the complex traits of interest. Therefore, it is important to emphasize that a gene module approach is by no means a replacement for current single variant/gene approaches. The development of new and improved statistical methods that detect more causal genes, regardless of whether they have a direct or indirect effect on the trait, while reducing false positives ([101](#)), will only enhance other approaches that rely on them, such as the gene module-based methods described here. In this review, we demonstrate that the identification of peripheral genes is key to prioritizing core genes that might represent more attractive drug targets, strongly suggesting that the mechanism proposed by the omnigenic model—i.e., that gene regulatory networks are highly interconnected and genes have specific roles with more direct or indirect effects on traits—is useful.

As the demand for analyzing large datasets and uncovering complex biological patterns grows, the need for faster and more efficient computational methods becomes increasingly critical. Future bioinformatics tools must not only handle the scale of these tasks but also ensure that they are accessible and adaptable across different computational environments. To meet these challenges, we are focusing on accelerating workflows with graphics processing unit (GPU) technology and exploring new platforms, such as WebAssembly, to bring powerful tools directly to users' browsers.

Leveraging GPU acceleration holds great promise for significantly speeding up bioinformatics workflows. Research has shown that GPU technology can provide remarkable performance boosts, with speed improvements reaching up to 1000x for specific tasks ([102](#)). Integrating GPU-aware libraries like CuPy ([103](#)) and RAPIDS ([104](#)) into our tools allows for seamless performance upgrades in established frameworks such as NumPy ([105](#)) and Scikit-learn ([106](#)). As our tools evolve, we anticipate developing custom GPU kernels to further enhance speed and optimize them for specific computational needs. Advanced GPU programming with PyCUDA ([107](#)) could also be key to future optimizations.

Another promising direction is utilizing WebAssembly ([108](#)) to make bioinformatics tools more portable and accessible. We aim to adapt existing bioinformatics applications into web-based tools that can be run directly in the browser, providing users with full control over their data and eliminating the need for data transfer over networks. Projects like Biowasm ([109](#)), Kana ([110](#)), and VirtualWasm ([111](#)) have already demonstrated the potential of WebAssembly for genomics and single-cell analysis, and we plan to build on this foundation. By executing statistical analyses, computations, and visualizations directly on the user's device, WebAssembly will not only address privacy and compliance concerns but also reduce server costs and computational load by shifting the workload to the client side.

Future innovations in GPU acceleration and WebAssembly will transform how bioinformatics tools are developed and deployed, leading to faster, more accessible solutions for researchers worldwide. By integrating these technologies, we can build the next generation of computational platforms that combine speed, flexibility, and privacy, unlocking new potential for biological discoveries. As these technologies mature, they will enable even more ambitious and large-scale analyses to be performed quickly and efficiently.